# Tuning Chemical Potential in the Dirac Cone by Compositional Engineering

R.K. Gopal, Sourabh Singh, Jit Sarkar, Chiranjib Mitra

Indian Institute of Science Education and Research Kolkata

Mohanpur - 741 246, India

We report the successful formation of bulk insulating ternary topological insulators candidate $Bi_2Se_2Te$ (BST) by pulsed laser deposition technique. The films were deposited with sequential ablation of separate $Bi_2Se_3$ (BS) and $Bi_2Te_3$ (BT) targets. From the X-ray diffraction analysis and temperature dependent resistivity we were able to conclude that the as grown thin films have ordered chalcogen layers and the chemical potential in these thin films lie in the bulk gap. To realize entirely topological transport for any device applications it is essential to tune the chemical potential in the bulk gap of the Dirac cone. Magnetotransport data exhibits pronounced two dimensional weak-antilocalization behavior (WAL) at low temperatures. BS and BT thin films do not exhibit topological transport as the chemical potential doesn't lie entirely in the bulk gap. It was found that BST thin films grown with this cost effective and simple yet elegant technique using double target can be used to deposit quaternary TI thin films, thereby tuning the chemical potential at will in the gap.

## Introduction:

Three dimensional topological insulators (TI) present unique state of quantum matter with insulating bulk and gapless spin polarized metallic surface states protected by Z2 topology of the bulk[1,2]. Linear dispersion between energy and momentum and large Fermi velocity of these surface electrons provide them the name Dirac electrons. These surface electronic states form a Dirac cone in the bulk band gap of these materials similar to graphene[3,4]. Experimental techniques probing these surface states such as angle (spin) resolved photoemission spectroscopy, scanning tunneling microscopy and spectroscopy as well as electrical measurement techniques have unequivocally shown signature of surface Dirac fermions till now. Due to nontrivial topology of the bulk band structure and spin momentum locking due to high spin-orbit interaction these materials could revolutionize conventional electronic devices. Quantum Hall effect in three dimensions TI has recently been observed in these materials after a long period of time of 25 years after its two dimensional counterpart had been discovered[5].

To observe the clear signature of the Dirac fermions and associated exotic phenomena one needs a highly insulating bulk with zero or negligible contribution to conductivity from bulk electrons and a super metallic surface conduction. But it was not possible with the second generation bismuth or antimony based TI's such as $Bi_2Se_3$ (BS), $Bi_2Te_3$ (BT) and $Sb_2Te_3$ (ST). These TI's were found to be heavily doped owing to the antisite defects and Se vacancies, hence a poor insulator in the bulk. Recently new TI's have been discovered showing high bulk insulating behavior, thus the focus from the binary three dimensional topological insulators BS, BT and ST is shifting because of a large bulk contribution to the conduction[6–8]. This in turn downplays the surface Dirac nature in these materials, the main hurdle in the development of these materials as noble technological applications such as low power spintronics. Compensation doping in the bulk and application of gate voltage to locate the Fermi level in the bulk band gap have been used as a remedy to get rid of the afore mentioned problems with BS and BT. But these solutions do not provide the ultimate goal of a perfect TI material since doping leads to further structural defects and disorder and extrinsic tuning of chemical potential is limited for specific substrates[9].

It was proposed theoretically and observed experimentally soon afterwards that alloying any two binary TI materials (BS & BT or BS & ST) provides a better TI material with a high bulk insulating character and dominant surface state transport. An insulating bulk is ensured by the location of the Fermi level in the bulk band gap. An ideal TI must have isolated Dirac cone, lesser hexagonal warping (lesser deformed Dirac cone), better spin texture and minimal contribution to bulk conduction which results in super metallic surface Dirac states. $Bi_2Se_2Te$ (BST), $Bi_2Te_2Se$ (BTS) and BiSbTeSe (BSTS) are the three remarkable bulk insulating Dirac materials which have shown surface dominated transport in terms of robust quantum interference phenomena: Weak Antilocalization (WAL) and quantum oscillations originated solely from surface Dirac states[10,11]. Remarkably, the alloying of BS, ST and BT was discovered to be more insulating in the bulk which means displaying a clear signature of surface Dirac fermions in transport properties and other surface sensitive techniques[12].

The purpose of this work is to find an ideal TI with a simpler surface electronic structure and find an easier synthesis procedure which is both cost effective and also provides us with the ability to tune the Fermi level in the middle of the band gap. Here we have deposited one of the most promising and bulk insulating ternary tetradymite BST thin films using separate BS & BT targets by pulsed laser deposition technique. Detailed morphological and structural analyses of these polycrystalline thin films were carried out by scanning electron microscope, X- ray diffraction, energy dispersive x- ray dispersion and Raman spectroscopy. Alongside BST we have also grown BTS thin films using the same procedure but the as grown films were found to be segregated into primarily BT phase and marginal Se clusters. While slightly Se rich form of BST ($Bi_2Se_{2.3}Te_{0.6}$) which is more ordered and designated as PS3, shows bulk insulating character and dominant surface transport at low temperatures by Dirac states, the stoichiometric BST ($B_2Se_2Te$) which is designated as PS2, exhibits metallic behavior as can be seen from the transport data. From the temperature dependent behavior of resistance it is possible to see that, below a certain temperature the robust character of the surface Dirac fermions prevail over disorder and electron – phonon scattering . As a result resistance of the film decreases with temperature below 45K, which is in contrast to the semiconducting or complete insulating behavior. This behavior is a manifestation of the topologically protected nature of Dirac fermions such that there is no backscattering and the surface electrons are robust against the usual scattering effects observed in other trivial two dimensional systems. This is termed as weak antilocalization. The onset of antilocalization below a certain temperature is due to locking of spin degrees of freedom of the electrons with their momentum and an associated "π "Berry phase of Dirac fermions in a bulk insulating TI.

## Experimental:

A complete set of thin films at constant optimized temperature of 300°C, base pressure $6.00 \times 10^{-1}$ mbar and laser energy density of $1.5 J/cm^2$ were deposited by varying the number of laser shots to ablate both the targets sequentially. Substrate temperature is also an important parameter for film growth. The composition and quality of the grown thin film is affected by the reaction rate and kinetics of the species. The substrate temperature controls these two mechanisms during thin film growth. Thus the optimization of the substrate temperature is a key factor. When it is too low, adatoms will not have adequate amount of energy to occupy the lowest potential site. And when the deposition temperature is on the higher side it results in three dimensional island like structures and non-uniform grains. Total number of laser shots was kept fixed at 50 for each cycle for both of the thin films and were varied in number such as 30:20, 25:25, 20:30, 40:10 (BS: BT) starting from the individual targets of BS and BT. As grown thin films were annealed in Argon environment (8mbar) at temperatures lower than the deposition temperature (300°C) as well as higher than the deposition temperatures and then cooled in controlled manner down to room temperature in argon atmosphere.

The substrates were cleaned in a well-defined manner in order to avoid surface contamination by carbon and oxygen. Therefore prior to deposition substrates were put in ultrasonic bath in the following sequence – 20 minutes in isopropyl alcohol, 5 minutes in distilled (DI) water, 20 minutes in acetone, 5 minutes in DI water, 5 minutes in DI water, 5 minutes in 5% hydrogen fluoride and DI water and finally in pure DI water for 10 minutes. Using this procedure we were able to achieve a smooth and clean substrate surface for better deposition.

All the substrates were degassed in the vacuum before deposition at a temperature of 350°C for 1hr which further ensures lesser contamination of the substrate surface. The target to substrate distance was kept at 5cm to confirm that the laser ablated plume could reach the substrates properly. The repetition rate of laser pulses was kept fixed at 2Hz in all the depositions. Prior to the growth of the films base pressure in the chamber was optimized at a value of $6.0 – 6.3 \times 10^{-1}$ mbar in order to achieve a good plume which converges properly on the substrates. Thickness of these films was found to be in the range of 200nm.

# Characterization:

From the SEM images it was found that the first film, PS2, in the series exhibited mainly triangular like structures structure as shown in the Figure 1. This is a trademark surface structure of the TI thin films and is consistent in structure with the previous reports on these alloys. The chemical composition of this film was found to be nearly stoichiometric $Bi_2Se_2Te$ by EDX. It is remarkable that these thin films were uniform in their structure and composition with no cluster formation (figure 1) which is the result of intermediate temperature annealing. The second thin film PS3 with composition $Bi_2Se_{2.3}Te_{0.6}$ also exhibited different hexagonal type flake like structure with larger and uniform grain size as compared to PS2 as shown in SEM pictures in the figure1. It is the larger grain size, homogeneity and uniform nature of thin film PS3 which is responsible for ordered crystal structure and insulating character.

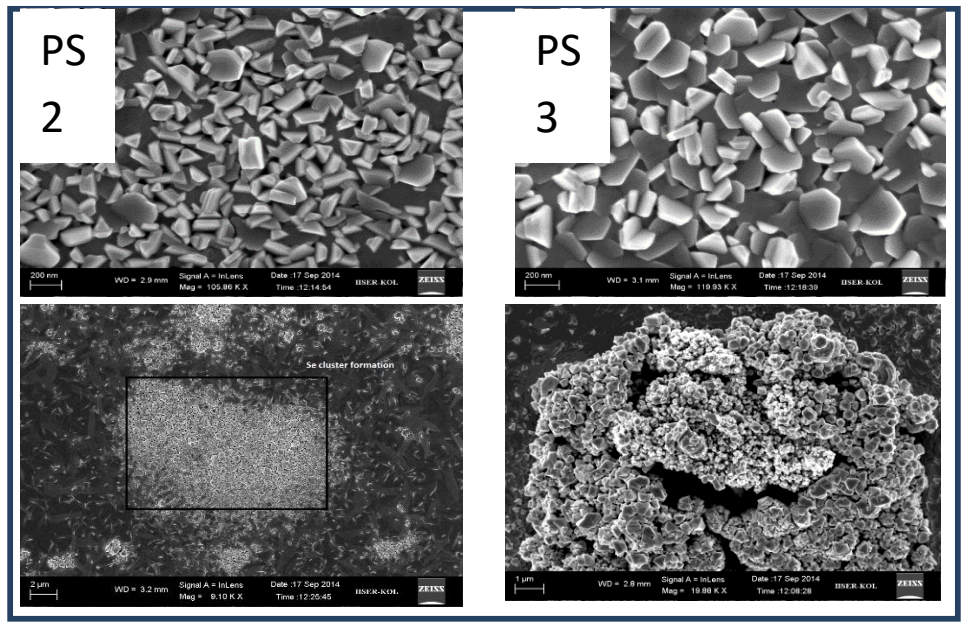

FIG. 1 SEM image of the thin films.

Other three films of this series PS were nearly BT in nature which is evident from the Raman characteristic peaks, and formed non-uniform surface structure with the formation of the many clusters and particulates of Se atoms at surface. It consisted of large clusters as seen from SEM images. We can attribute this cluster formation to the segregation of selenium atoms at different spatial location. This can be seen from the designated rectangular portion in fig. 1(c).

Similar sequences of the deposition were performed to deposit BTS thin films. These films too were deposited at 300°C followed by annealing at 250°C for two hours and then subsequently cooled to room temperature in about 2 hours. As seen from the SEM images here too there are a number of cluster formations on the surface with no flake like structure characteristic of TI thin films. These thin films show Raman peaks corresponding to BT which suggests that low temperature annealing of as grown thin films is not energetically favorable for BTS phase formation ( at least deposited on 300°C) and the dominant phase remains that of BT in the films suggesting disordered phase formation. BTS has a more ordered structure because of the central-layer substitution while BST on the other hand, has a somewhat disordered structure which is due to random Te substitutions of Se atoms in the outer QLs. We believe that this might be the reason that we are not able to get BTS thin films using this technique.

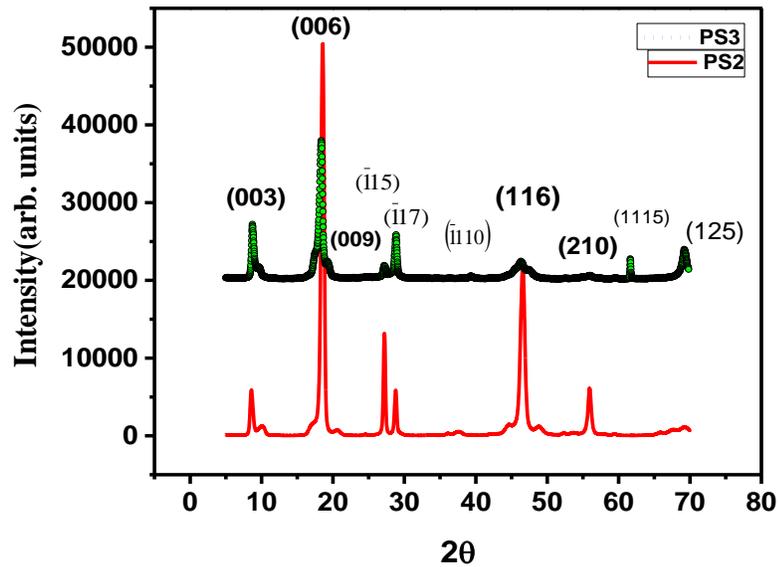

FIG. 2. XRD data of PS2 and PS3 samples shown together. While Intensity of peaks in PS2 is higher than that of PS3 consistent with the Raman spectra, some of the peaks are missing in PS2. XRD pattern of PS3 shows deviations from the exact that of ordered skippenite structure which suggests disordered occupation of Se/Te atoms in the crystal lattice.

**Raman and XRD Analysis:**

In order to determine crystal structure and crystalline phase information Raman spectra and XRD data of as grown thin films were taken. Raman spectrum clearly resolves three peaks corresponding to the 15 vibration modes (12 optical & 3 acoustic), two doubly degenerate $E_{2g}$ (in plane vibration modes) and $A^2_{1g}$ (out of plane vibrations), characteristic to BST phase which is consistent with previous reports[13]. Ternary tetradymites, like binary TI materials possess rhombohedral crystal structure (space group $R\bar{3}m$) with layered hexagonal packing in the c-axis direction. Both of the compositions, Se rich (PS3) and stoichiometric (PS2) thin films exhibit Raman active modes corresponding to the Se rich phases of BST. The intensity of all three modes in case of PS2 is much larger than that of PS3, which indicates that some of the bonds in the former have increased. On close examination the Raman peaks in case of PS3 (which has been found to be of insulating character, see fig.4), are red shifted and broader with respect to that of PS2. Though these two films have different compositions and different grain sizes and comparing the Raman peak shifts and intensities may not be entirely relevant, however, the peak positions of all Raman modes for both are nearly the same. This reduced intensity, and marginal shift in Raman peaks coupled with uniform grains (in PS3) may be responsible for the different temperature dependent behavior of resistivity (fig. 4). However, further fine tuning of composition is required to establish this connection and a detailed study is in progress to ascertain this. It should be noted that the insulating nature of these TI materials is dependent on the compensation of the accepter vacancies by the donor atoms, which in this case are Te atoms occupying Se vacancies, which have lower Se/Te defect formation energy in BST. Owing to this uncompensated nature and disordered occupation of the Se/Te site it shows much lesser bulk resistivity at low temperatures in comparison to BTS which has an ordered structure.

XRD data taken on these films is compared with the previously reported data and indicates the crystalline nature of as grown thin films. While very sharp peaks with high intensity and lesser full width at half maxima (FWHM) are seen in PS2, the XRD pattern of PS3 shows broader peaks with lesser intensity as shown in fig.2. XRD analysis shows that PS3 also has nearly skippenite structure with slight deviation from the ordered structure, with space group $R\bar{3}m$. Some of the peaks in the PS2 are missing as compared to PS3 which suggests that PS2 has larger deviation from the ordered skippenite structure as compared to PS3.

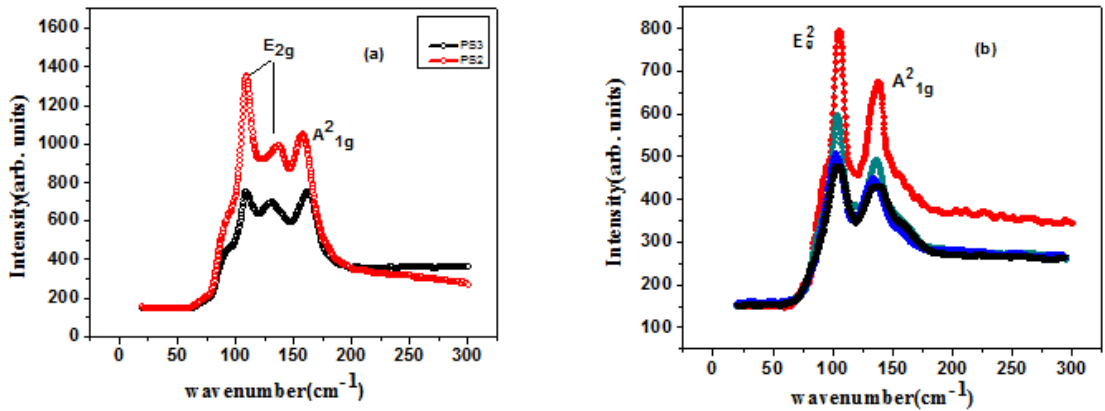

FIG3. (a) Raman spectrum of PS2 and PS3 samples. The locations of the peak positions for PS2 (109.42 cm$^{-1}$, 137.22 cm$^{-1}$ and 157.32 cm$^{-1}$) while for PS3 (109.09 cm$^{-1}$, 131.19 cm$^{-1}$, 162.34 cm$^{-1}$). Raman spectrum of the as grown BST and BT thin films clearly show the characteristic peaks corresponding to the BST vibration modes, doubly degenerate $E_{2g}$ and $A^2_{1g}$. (b) While other four thin films exhibits Raman modes characteristic to BT phase with $E^2_g$ and $A^2_{1g}$ centered around 105.03 and 137.53 respectively.

## **Transport Properties:**

After characterizing our thin films with all the necessary tools, we focus our attention towards the transport properties of the BST thin films. The R-T nature of our sample (PS3) shows an activated behavior down to 50K and thereafter it starts to drop in the lower temperature regime. In an earlier work we had reported bulk insulating behavior in BST thin films with different composition but Se rich[14]. But it differs from the present work since those samples were prepared by using a single BST target.

From the temperature dependent resistance/resistivity (R - T) plot in fig.3 we can see BST thin films exhibit insulating character down to 45K, while below this temperature it shows bulk insulating to two dimensional Dirac metal transition which is the signature of topological delocalization of TI surface states owing to reduced Dirac electron – phonon scattering. This transition has been reported in other bulk insulating TI thin films before. This transition is a manifestation of the robustness of 2D Dirac fermion towards localization by disorder and electron-electron interactions as soon as one goes below 45K. Therefore one can observe topological nature of surface states by the behavior of the R-T in a bulk insulating TI samples, owing to robust Z2 delocalization and "π" Berry phase[14].Topological protection is a combined effect of Z2 topology of the bulk electronic structure and "π" Berry phase, which is absent in graphene. Due to this fundamental difference graphene is not a topological insulator and exhibit and weak localization (WL) due to presence of disorder. Even in the high mobility graphene samples one can observe WAL in the low field and WL in the high field regime a crossover**.**. This is due to presence of an even number of Dirac cones which in turn results in the inter-valley scattering manifested in MR data.

On the other hand the R – T data of the sample PS2 displays a complete metallic character down to 25K and then shows an insulating ground state due to the competing effects of freezing of the bulk electronic states and electron-electron interaction. Therefore the chemical potential of PS2 thin film lies in the bulk conduction band which is consistent with the ARPES studies and recent transport studies on single crystals of this composition[7,15,16]. This behavior is similar to the previously deposited metallic binary and ternary thin films, where resistance makes an upturn at low temperatures[17]. From the combined observation of the R – T data, XRD pattern, Raman peaks, SEM images and EDX spectra we conclude that the completely stoichiometric BST thin films ($Bi_2Se_2Te$) are metallic and exhibit disordered and random occupation of Se/Te atoms in the quintuple layer, while Se rich BST thin films ($Bi_2Se_{2.3}Te_{0.7}$) show near ordered occupation in the quintuple layer which results in bulk insulating character, thus pushing the chemical potential in the bulk band gap. It has been observed previously in other studies that non-stoichiometric BST thin films ($Bi_2Se_{1.7}Te_{1.3}$) annealed at slightly elevated temperatures (Deposition – 250, Annealed - 300) for about an hour leads to insulating character. While larger compensation is observed in PS3 by Te/Se atoms instead of forming bonds with Bi or Te/Se atoms in the quintuple layer unit, PS2 remains uncompensated which is reflected in the temperature dependent resistivity measurement displaying a metallic profile (fig.4). Therefore the activated behavior of the PS3 is largely due to the disordered and random occupation of Se vacancies by Te atoms and the overall crystal structure remains disordered and the doped acceptors get somewhat compensated.

Magnetoresistance data of TI materials provide a plethora of information about the characteristic transport parameters such as phase coherence length $l\varphi$, elastic mean free path $le$, mobility $\mu$ and the phase associated with the Dirac fermions. Magnetotransport measurement can therefore throw light on various phenomena like WAL (at low fields), LMR (at high fields) and SdH oscillations for samples with very high mobility. Our PLD grown thin films do not yield a very high value of mobility ($25 cm^2/V$-sec) therefore are in the diffusive transport regime. By diffusive transport we mean $l_{so} < l_e < l_\varphi < L$, where $l_{so}$, $l_e$ and $l_\varphi$ are the characteristic length scales corresponding to spin-orbit scattering, elastic scattering and inelastic scattering[18,19]. In order to extract transport parameters from the MR measured on our BST thin films we fit Hikami, Larkin and Nagaoka (HLN) expression for 2D electron systems with strong spin orbit interaction[20].

$$\Delta G(B) = \frac{e^2}{2\pi h}\left[\psi\left(\frac{B_\varphi}{B}+\frac{1}{2}\right) - \ln\left(\frac{B_\varphi}{B}\right)\right] - \frac{e^2}{\pi h}\left[\psi\left(\frac{B_{SO}+B_e}{B}+\frac{1}{2}\right) - \ln\left(\frac{B_{SO}+B_e}{B}+\frac{1}{2}\right)\right] + \frac{3e^2}{2\pi h}\left[\psi\left(\frac{\left(\frac{4}{3}\right)B_{SO}+B_\varphi}{B}+\frac{1}{2}\right) - \ln\left[\left(\frac{\left(\frac{4}{3}\right)B_{SO}+B_\varphi}{B}\right)\right]\right]$$

For $B \ll B_e$ and $B \ll B_\varphi$

$$\boldsymbol{\delta G_{WAL}(B) \equiv G(B) - G(0) \cong \alpha\frac{e^2}{2\pi^2 \hbar}\left[\Psi\left(\frac{1}{2}+\frac{B_\varphi}{B}\right) - \ln\left(\frac{B_\varphi}{B}\right)\right]} \qquad (1)$$

Where $\Psi$ is digamma function and $B_\varphi = \hbar/4el\varphi^2$ is characteristic magnetic field corresponding to the coherence length $l_\varphi$. At higher magnetic fields the above used approximations do not hold, moreover we must include classical MR terms also at higher field values. So, Assaf et al. obtained a modified HLN equation which comprises of a quadratic magnetic field contributing term with a coefficient $\beta$ which consists of both classical and quantum parts[21].

$$\Delta G(B) = \frac{\alpha e^2}{2\pi h}\left[\psi\left(\frac{\hbar}{4el_\varphi^2 B} + \frac{1}{2}\right) - \ln\left(\frac{\hbar}{4el_\varphi^2 B}\right)\right] + \beta B^2 \qquad (2)$$

The value of coefficient "α" provides the information about the effective number of coherent channels contributing to conduction. Alternatively the transport coefficient "α" has its importance in determining various transport properties of TI thin films/devices depending on the behavior of the resistivity with temperature i.e., either metallic or bulk insulating. In the metallic thin films such as binary TI's the value of α generally have been found to be approximately -0.5. Earlier due to lack of experimental evidences it was suggested that only top surface contributed to the conduction while bottom one did not owing to large lattice mismatch and disorder present at the substrate – thin film interface. But recent experimental observations of two surface quantized conduction in the 3D intrinsic half integer quantum Hall effect in the bulk insulating quaternary and ternary TI's have confirmed that both top and bottom surfaces conduct[3]. Therefore the value for the metallic thin films must lie in the range of 0.5 – 0.6 since conduction is bulk mediated, and surface states couple with the bulk, stated to make an effective single 2D channel.

While in case of bulk insulating TI thin films/devices one should get α near to -1 since bulk channels do not contribute in the transport and it has been verified experimentally in many reports where the samples were not exposed to the ambient environment. By fitting HLN equation to the quantum correction to the conductance the value of α have been extracted as -0.57 which is consistent with the values reported earlier on TI thin films. The value of α must be -1 corresponding to the bottom and top surface channels but the value extracted from the fit is lesser than the expected one. There may be two possible reasons for this discrepancy. The reason for this reduced value of α lies in the coexistence of trivial two dimensional electron gas along with Dirac surface states on the top surface owing to band banding near surface. These 2D bands come in to effect when TI sample is exposed to the external ambient environment as is experimentally evidenced by many ARPES measurements[22,23]. Well defined two or more parabolic 2D bulk sub bands located just below bulk conduction band have been observed in these studies. This topologically trivial 2D electron gas may be in the localization regime. While 2D electron gas at the top surface makes a positive contribution to the transport coefficient α by having a localization effect, the Dirac gas contributes from both the surfaces negatively. Another reason for reduced value of α can be the larger disorder created at the bottom surface due to lattice mismatch in the BST thin films and Si (100) substrates, driving the bottom surface in to highly disordered interface [9,24]. Therefore the current flows on the lesser resistive path which is naturally the upper surface as compared with the bottom one, resulting in the value of α close to -0.5.

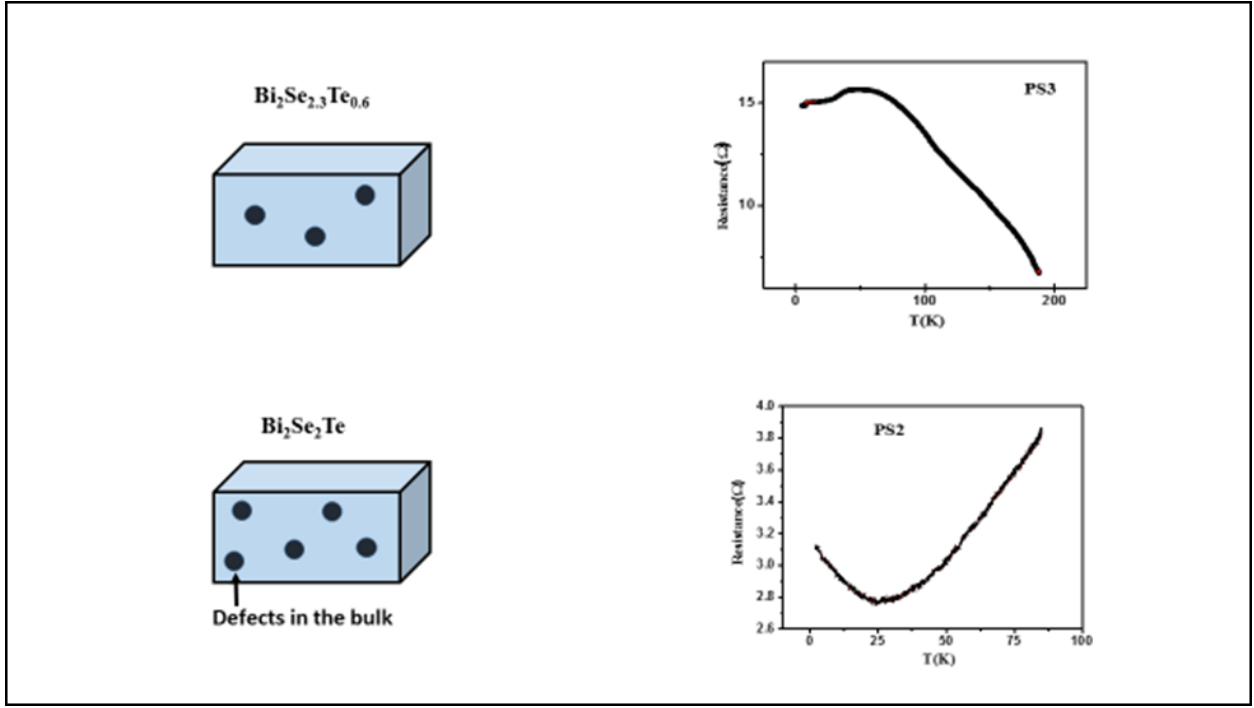

FIG. **4** Resistance vs. Temperature behavior of PS2 and PS3 sample. The left panel depicts that excess of Se quenches the defects in the bulk thus resulting in bulk insulating behavior.

In order to probe quantum interference phenomena in TI, MR measurements at high fields were conducted at two different temperatures shown in the fig.5 (a). BST thin films respond to the perpendicular magnetic field as sharp logarithmic increase in the MR around the zero magnetic field, as governed by the HLN eqn. a characteristic feature of TI thin films.

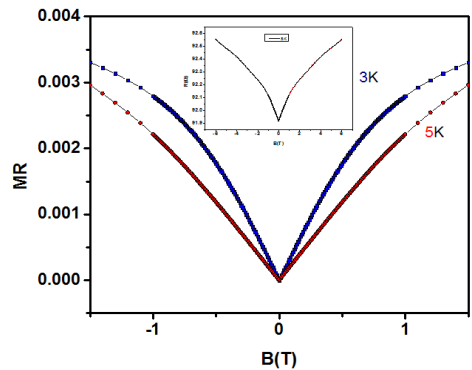

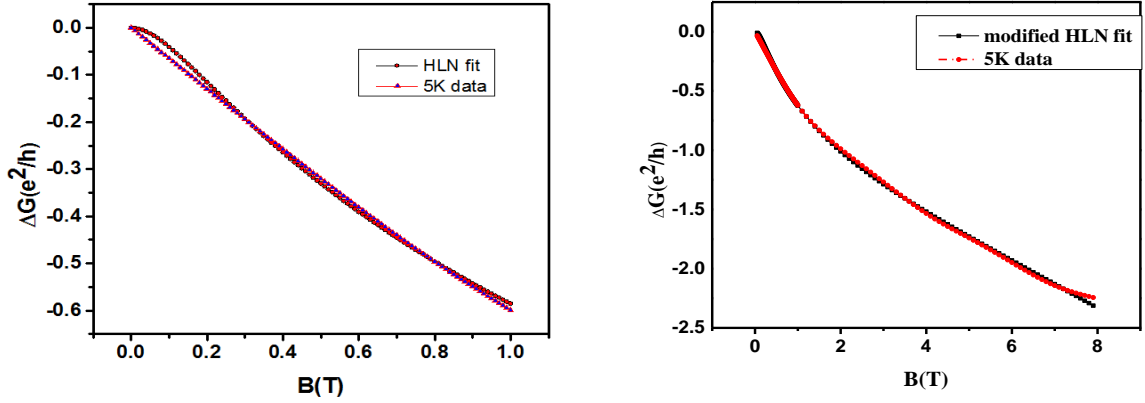

FIG.5 **(a)** Weak antilocalization effect seen for PS3 sample at two different temperatures. The inset shows the MR behavior till 8T. (b) Magnetoconductance data for 5K fitted with HLN equation. The fit yields α=0.57 and $l_\Phi = 48.6\ nm$. (c) fitting of high field magnetoconductance with modified HLN equation with classical term added to HLN equation.

Although BST has been predicted to have a disordered occupation of Se/Te atoms in the QL of the crystal structure of TI's, observed insulating character though raises question on these claims. The ternary tetradymite TI materials are very sensitive to the Se/Te ratio (Specially in BST) which may drive them as metallic or insulating depending on the ordered occupation in the QL such as BTS, similar arguments hold for BST as well since it is found metallic and insulating on changing Se/Te ratio slightly[12,14,21,25,26]. Therefore such a systematic study was called for and hence carried out.

## *Conclusion:*

In conclusion, we have successfully grown high quality BST thin films by using double target technique. But BTS phase with this technique could not be grown. But we suggest BTS phase can be achieved by further decreasing the number of shots in one sequence of ablation i.e. BT: BS – 5:2, 7:3 or little higher. BST thin films show weak antilocalization behavior at low temperatures, which is fitted with HLN equation and the modified HLN equation at low and high magnetic fields respectively. The obtained fit parameters "α" and phase coherence length are consistent with the thin films grown with other techniques. We claim that, with further fine tuning the chemical composition in between BS, BT and ST ($Sb_2Te_3$) this technique can provide new materials unknown so far to the TI working fraternity, which are not only more insulating in the bulk but also have a simpler surface electronic structure. This tailoring of the bulk using PLD technique is economical, fast and controllable. The ability to manipulate and tune the chemical potential in topological insulator thin films paves the path for noble applications and multifunctional devices in the spintronics sector.

**Acknowledgement:**

The authors would like to thank Ministry of Human Research Development (MHRD) and IISER Kolkata for the financial assistance. SS and JS would like to thank University Grants Commission (UGC) for funding. Authors would like to thank Dr. G.D. Mukherjee and Rajesh Jana for the Raman measurements. RKG is grateful to Dr. Bhavesh Patel and Dr. MK Murary for the helpful discussion.